\documentclass[12pt]{article}

\begin{document}
\begin{center}
 {\bf{\Large Black Hole in a Model with Dilaton and Monopole Fields II}}
 \vspace{1cm}

E. Kyriakopoulos\footnote{E-mail: kyriakop@central.ntua.gr,
 Tel: 00302107722980, Fax: 00302107722932}\\
Department of Physics\\
National Technical University\\
157 80 Zografou, Athens, GREECE
\end{center}

\begin {abstract}
We present an exact black hole solution in a model having besides
gravity a dilaton and a monopole field, which is a generalization
of a black hole solution we have found. The new solution, as the
previous one, has three free parameters, one of which can be
identified with the monopole charge, and another with the ADM
mass. Its metric is asymptotically flat, has two horizon,
irremovable singularity only at $r=0$, and the dilaton field is
singular only at $r=0$. The dominant and the strong energy
condition are satisfied outside and on the external horizon.
According to a formulation of the no hair conjecture the solution
is "hairy". Also a reformulation of the model with two monopole
fields is given, which results in the appearance of an additional
symmetry and therefore in the appearance of a conserved dilaton
charge.

PACS number(s): 04.20.Jb, 04.70.Bw, 04.20.Dw

\end {abstract}

 According to a formulation of the no hair
conjecture \cite{Nu} :"We say that in a given theory there is
black hole hair when the space-time metric and the configuration
of the other fields of a stationary black hole solution are not
completely specified by the conserved charges defined at
asymptotic infinity". Previously \cite{Ky} in a four-dimensional
model having besides gravity a dilaton and a monopole field we
found a solution which is "hairy" according to the above
formulation. Other formulations of the no hair conjecture can also
be found in the literature \cite{Wh}-\cite{Bi}. In this work we
shall generalize our previous model. Also we shall reformulate the
model such that an additional symmetry appears and therefore a
conserved dilaton charge can be defined.

 Consider the action
\begin{equation}
\int d^{4}x \sqrt{-g} L = \int d^{4}x \sqrt{-g}
\{R-\frac{1}{2}\partial_{\mu}\psi\partial^{\mu}\psi-
f(\psi)F_{\mu\nu}F^{\mu\nu}\} \label{1}
\end{equation}
\begin{equation}
f(\psi)=g_{1}e^{(c+\sqrt{c^2+1})\psi}
+g_{2}e^{(c-\sqrt{c^2+1})\psi} \label{2}
\end{equation}
where $R$ is the Ricci scalar, $\psi$ is a dilaton field, $c$,
$g_1$ and $g_2$ are real constants and $F_{\mu\nu}$ is a pure
monopole field
\begin{eqnarray}F=Q sin\theta d\theta\wedge d\phi \label{3}
\end{eqnarray}
with $Q$ its magnetic charge.From this action we find the
following equations of motion
\begin{equation}
(\partial^{\rho}\psi)_{;\rho}-\frac{df}{d\psi}F_{\mu\nu}F^{\mu\nu}=0
\label{4}\end{equation}
\begin{equation}
(f F^{\mu\nu})_{;\mu}=0 \label{5}
\end{equation}
\begin{equation}R_{\mu\nu}=\frac{1}{2}\partial_{\mu}\psi\partial_{\nu}\psi+
2f(F_{\mu\sigma}{F_{\nu}}^{\sigma}-
\frac{1}{4}g_{\mu\nu}F_{\rho\sigma}F^{\rho\sigma})\label{6}
\end{equation}

We want to find static spherically symmetric solutions of the
above equations which are asymptotically flat and have regular
horizon. We write the metric in the form \cite{Ga}
\begin{equation}
ds^2=-\lambda^{2}dt^{2}+\lambda^{-2}dr^{2}+\xi^{2}d\Omega
\label{7}
\end{equation}
where $\lambda$ and $\xi$ are functions of $r$ only and
$d\Omega=d\theta^{2}+sin^{2}\theta {d{\phi}^{2}}$. From the above
metric and Eq (\ref{3}) we get
$F_{\mu\nu}F^{\mu\nu}=\frac{2Q^{2}}{\xi^{4}}$, and we can prove
that Eq (\ref{5}) is satisfied. The dilaton Eq (\ref{4}) for
$\psi=\psi(r)$ takes the form
\begin{equation}
(\lambda^{2}\xi^{2}\psi')'=2\frac{df}{d\psi}Q^{2}\xi^{-2}\label{8}
\end{equation}
where prime denotes differentiation with respect to $r$. The
non-vanishing components of the Ricci tensor of the metric
(\ref{7}) are $R_{00}$, $R_{11}$, $R_{22}$ and
$R_{33}=sin^{2}\theta{R_{22}}$, and for the first three components
we get respectively from Eqs (\ref{6}) the relations
\begin{equation}
(\lambda^{2})'' +(\lambda^{2})'(\xi^{2})'\xi^{-2}=2f Q^{2}\xi^{-4}
\label{9}
\end{equation}
\[-(\lambda^{2})''\lambda^{-2}-2(\xi^{2})''\xi^{-2}-
(\lambda^{2})'(\xi^{2})'\lambda^{-2}\xi^{-2}+
[(\xi^{2})']^{2}\xi^{-4}=(\psi')^{2}\]
\begin{equation}-2f Q^{2}\lambda^{-2}\xi^{-4} \label{10 }
\end{equation}
\begin{equation}
-[\lambda^{2}(\xi^{2})']'+2=2f Q^{2}\xi^{-2} \label{11}
\end{equation}
Eqs (\ref{8})-(\ref{11}) form a system of four equations for the
three unknowns $\lambda^{2}$, $\xi^{2}$ and $\psi$. We found the
following solution of this system
\begin{equation}
\lambda^{2}=\frac{(r+A)(r+B)}{r(r+\alpha)}(\frac{r}{r+\alpha})^\frac{c}{\sqrt{c^2
+1}},\>\>\>\>\xi^{2}=
r(r+\alpha)(\frac{r+\alpha}{r})^\frac{c}{\sqrt{c^2+1}}\label{12}
\end{equation}
\begin{equation}
e^{\psi}=e^{\psi_{0}}(1+\frac{\alpha}{r})^\frac{1}{\sqrt{c^2+1}}
\label{13}
\end{equation}
where $A$, $B$, $\alpha$ and $\psi_{0}$ are integration constants,
provided that the following relations are satisfied
\begin{equation}
g_{1}=\frac{AB(\sqrt{c^2+1}-c)}{2Q^{2}\sqrt{c^2+1}}e^{-(c+\sqrt{c^2+1})\psi_{0}}
\label{14}
\end{equation}
\begin{equation}
 g_{2}=\frac{(\alpha-A)(\alpha-B)(\sqrt{c^2+1}
+c)}{2Q^{2}\sqrt{c^2+1}}e^{-(c-\sqrt{c^2+1})\psi_{0}} \label{14'}
\end{equation}
From Eqs (\ref{7}) and (\ref{12}) we get
\[ds^{2}=-\frac{(r+A)(r+B)}{r(r+\alpha)}(\frac{r}{r+\alpha})^\frac{c}{\sqrt{c^2
+1}}dt^{2}+\frac{r(r+\alpha)}{(r+A)(r+B)}(\frac{r+\alpha}{r})^\frac{c}{\sqrt{c^2
+1}}dr^{2}\]
\begin{equation}
+r(r+\alpha)(\frac{r+\alpha}{r})^\frac{c}{\sqrt{c^2+1}}d{\Omega}
\label{15}
\end{equation}
Our solution is given by Eqs (\ref{3}) and (\ref{13})-(\ref{15}).

From Eq. (\ref{15}) we get asymptotically
\begin{equation}
-g_{00}=1-\frac{\alpha(1+\frac{c}{\sqrt{c^2+1}})-A-B}{r} +
O(r^{-2})\label{16}
\end{equation}
 Therefore the solution
is asymptotically flat and its Arnowitt-Deser-Misner (ADM) mass
$M$ is given by
\begin{equation}
2M=\alpha(1+\frac{c}{\sqrt{c^2+1}})-A-B \label{17}
\end{equation}

It is obvious that $\psi_{0}$ is the asymptotic value of $\psi$.
Also for the choice
\begin{equation}
\alpha>0,\>\>\>\>\> A<0,\>\>\>\>\> B<0 ,\label{18}
\end{equation}
we have
\begin{equation}
g_{1}>0,\>\>\>\>\> g_{2}>0,\>\>\>\>\> M>0,\label{19}
\end{equation}
and $\psi$ is singular only at $r=0$. We shall make this choice
for $\alpha$, $A$ and $B$.

The solution has the integration constants $\alpha$, $A$, $B$,
$\psi_{0}$ and $Q$, which for given $g_{1}$, $g_{2}$ and $c$ must
satisfy Eqs (\ref{14}) and (\ref{14'}). Therefore only three of
them are independent. Introducing the ADM mass $M$ by the relation
(\ref{17}) we can take $M$, $Q$ and $\psi_{0}$ as independent
parameters and express the parameters $\alpha$, $A$ and $B$ of the
dilaton field and the metric in terms of then. This means that the
solution has arbitrary mass, arbitrary magnetic charge and an
additional arbitrary parameter. Therefore it is a "hairy" solution
according to the definition given in Ref. \cite{Nu}. We should
point out that the well known GHS-GM solution of D. Garfinkle, G.
T. Horowitz and A. Strominger  \cite{Ga}, found previously by G.
W. Gibbons and by G. W. Gibbons and K. Maeda \cite{Gi} is not a
"hairy" solution according to this definition \cite{Su}.

The metric coefficient $g_{rr}$ is singular at $r=-A$ and  $r=-B$
while the coefficient $g_{tt}$ is singular at $r=0$ but not at
$r=-\alpha$, since $\alpha>o$. However the singularities at $r=-A$
and $r=-B$ are coordinate singularities. Indeed the Ricci scalar
$R$ and the curvature scalar
$R_{\mu\nu\rho\sigma}R^{\mu\nu\rho\sigma}$ are given by
\begin{equation}
R=\frac{\alpha^{2}(r+A)(r+B)}{2(c^2+1)r^{3}(r
+\alpha)^{3}}(\frac{r}{r+\alpha})^\frac{c}{\sqrt{c^2+1}}\label{20}
\end{equation}
\begin{equation}
R_{\mu\nu\rho\sigma}R^{\mu\nu\rho\sigma}=
\frac{P(r,\alpha,A,B)}{4(c^2+1)^2
r^{6}(r+\alpha)^{6}}(\frac{r}{r+\alpha})^\frac{2c}{\sqrt{c^2+1}}
\label{21}
\end{equation}
where $P(r,\alpha,A,B)$ is a complicated polynomial of $r$,
$\alpha$ $A$ and $B$. Therefore $R$ and
$R_{\mu\nu\rho\sigma}R^{\mu\nu\rho\sigma}$ are regular at $r=-A$
and $r=-B$ , which means that only at $r=0$ we have a real, an
irremovable  singularity.

If we make the Eddington-Finkelstein type transformation
\begin{equation}
dt=dt' \pm
\frac{(A+B)r'+AB}{(r'+A)(r'+B)}(\frac{r'+\alpha}{r'})^{1+\frac{c}{\sqrt{1+c^2}}}dr',\>\>r=r',\>\>\theta={\theta}',\>\>
\phi={\phi}'\label{22}
\end{equation}
the metric $ds^{2}$ of Eq. (\ref{15}) takes the regular at $r=-A$
and $r=-B$ form
\[ds^{2}=-\frac{(r'+A)(r'+B)}{r'(r'+\alpha)}(\frac{r'}{r'+\alpha})^\frac{c}
{\sqrt{c^2+1}}dt'^{2}+\frac{r'+\alpha}{r'^{3}}\{r'^{2}-(A+B)r'-AB\}\]
\begin{equation}
\times(\frac{r'+\alpha}{r'})^\frac{c} {\sqrt{c^2+1}}dr'^{2} \mp
\frac{2}{r'^{2}}\{(A+B)r'+AB\}dr'dt' +
r'(r'+\alpha)(\frac{r'+\alpha}{r'})^\frac{c} {\sqrt{c^2+1}}d\Omega
\label{23 }
\end{equation}
From the above expression we find that the radial null directions,
i.e. the directions for which $ds^{2}=d \theta=d \phi=0$, are
determined by the relations
\begin{equation}
r'(\frac{r'}{r'+\alpha})^\frac{c}
{\sqrt{c^2+1}}dt'\pm(r'+\alpha)dr'=0 \label{24}
\end{equation}
\begin{equation}
r'(r'+A)(r'+B)(\frac{r'}{r'+\alpha})^\frac{c} {\sqrt{c^2+1}}dt'
\mp (r'+\alpha)\{r'^{2}-(A+B)r'-AB\}dr'=0 \label{25}
\end{equation}
The solution of the above equations give the intersections of the
light cone with the $t'-r'$ plane. From the numerical solution of
these Eqs we find that the solution with the upper sign is a black
hole solution with two horizons at $r'=-A$ and at $r'=-B$, and the
solution with the lower sign is a white hole solution.

 The energy-momentum tensor $T_{\mu\nu}$ of our solution is
given by
\[T_{\mu\nu}=\partial_{\mu}\psi\partial_{\nu}\psi
+4f F_{\mu\rho}{F_{\nu}}^{\rho} -g_{\mu
\nu}\{\frac{1}{2}\partial_{\rho}\psi\partial^{\rho}\psi +f F_{\rho
\sigma}F^{\rho \sigma}\}\]
\[=\frac{\alpha^{2}}{(c^2+1)r^{2}(r+\alpha)^{2}}\delta_{\mu
r}\delta_{\nu
r}+\{\frac{2AB}{r^{2}}(1-\frac{c}{\sqrt{c^2+1}})+\frac{2(\alpha-A)(\alpha
-B)}{(r+\alpha)^{2}}\]\[\times(1+\frac{c}{\sqrt{c^2+1}})\}
(\delta_{\mu\theta}\delta_{\nu \theta} +{sin^{2}\theta}\delta_{\mu
\phi}\delta_{\nu \phi}) -g_{\mu \nu}\{
\frac{{\alpha^{2}}(r+A)(r+B)}{2(c^2+1)r^{3}(r+\alpha)^{3}}\]
\begin{equation}
+\frac{AB}{r^{3}(r+\alpha)}(1-\frac{c}{\sqrt{c^2+1}})
 +\frac{(\alpha-A)(\alpha-B)}{r(r+\alpha)^{3}}(1+\frac{c}{\sqrt{c^2+1}})
 \}(\frac{r}{r+\alpha})^\frac{c}{\sqrt{c^2+1}}
\label{26}
\end{equation}
Calculating the eigenvalues of $T_{\mu\nu}$ we can show that it
satisfies the dominant as well as the strong energy condition
outside and on the external horizon.

If $A<B<0$ and $\alpha>0$ the Hawking temperature $T_H$ of our
solution is \cite{Ho}
\begin{equation}
T_H=\frac{|{\lambda^2}'(-A)|}{4\pi} = \frac{(B-A)(-A)^{-1
+\frac{c}{\sqrt{c^2+1}}}(\alpha-A)^{-1-\frac{c}{\sqrt{c^2+1}}}}{4\pi}
\label{27}
\end{equation}

The previous model can be reformulated such that the Lagrangian of
the new model has an additional symmetry. To do that let as
consider the action
\begin{equation}
\int d^{4}x \sqrt{-g} L = \int d^{4}x \sqrt{-g}
\{R-\frac{1}{2}\partial_{\mu}\psi\partial^{\mu}\psi-
h(\psi)F_{\mu\nu}F^{\mu\nu} -k(\psi)G_{\mu\nu}G^{\mu\nu}\}
\label{28}
\end{equation}
\begin{equation}
h(\psi)=g_{1}e^{(c+\sqrt{c^2+1})\psi},\>\>\>\>\>
k(\psi)=g_{2}e^{(c-\sqrt{c^2+1})\psi}\label{29}
\end{equation}
where $R$ is the Ricci scalar, $\psi$ is a dilaton field, $c$,
$g_1$ and $g_2$ are real constants and $F_{\mu\nu}$ and
$G_{\mu\nu}$ are two pure monopole fields
\begin{eqnarray}F=Q_{1}sin\theta d\theta\wedge d\phi,\>\>\>\>\> G=Q_{2}sin\theta
d\theta\wedge d\phi  \label{30}
\end{eqnarray}
where $Q_{1}$ and $Q_{2}$ are their magnetic charges respectively.
The above action for $g_{1}=g_{2}=1$ and $c=0$ is the part of the
$SO(4)$ version of the $N=4$, $d=4$ supergravity action
$I_{SO(4)}$ without the axion field \cite{Ka} . Also the above
action is invariant under the transformations
\begin{eqnarray}
g_{\mu\nu} \rightarrow g_{\mu\nu},\>\>\psi \rightarrow \psi+b,
\>\>\>F_{\mu\nu} \rightarrow
e^{-(\frac{c+\sqrt{c^2+1}}{2})b}F_{\mu\nu},\>\>G_{\mu\nu}
\rightarrow e^{-(\frac{c-\sqrt{c^2+1}}{2})b}G_{\mu\nu} \label{31}
\end{eqnarray}
where $b$ is a constant. From this symmetry the conserved dilaton
current
\begin{equation}
j_\mu=\partial_{\mu}\psi+2(c+\sqrt{c^2+1})h(\psi)F_{\mu\nu}A^\nu+
2(c-\sqrt{c^2+1})k(\psi)G_{\mu\nu}B^\nu\label{31'}
\end{equation}
is obtained and a conserved dilaton charge can be defined.

 From the above action the following equations of motion
\begin{equation}
(\partial^{\rho}\psi)_{;\rho}-\frac{dh(\psi)}{d\psi}F_{\mu\nu}F^{\mu\nu}-
\frac{dk(\psi)}{d\psi}G_{\mu\nu}G^{\mu\nu}=0
\label{32}\end{equation}
\begin{equation}
\{h(\psi)F^{\mu\nu}\}_{;\mu}=0,\>\>\>\>\{k(\psi)G^{\mu\nu}\}_{;\mu}=0
\label{33}
\end{equation}
\[R_{\mu\nu}=\frac{1}{2}\partial_{\mu}\psi\partial_{\nu}\psi+
2h(\psi)(F_{\mu\sigma}{F_{\nu}}^{\sigma}-\frac{1}{4}g_{\mu\nu}F_{\rho\sigma}F^{\rho\sigma})\]
\begin{equation}+2k(\psi)(G_{\mu\sigma}{G_{\nu}}^{\sigma}-
\frac{1}{4}g_{\mu\nu}G_{\rho\sigma}G^{\rho\sigma})\label{34}
\end{equation}
are obtained. To get static spherically symmetric solutions of the
these equations which are asymptotically flat and have regular
horizon we write the metric in the form of Eq. (\ref{7}). Then we
get from Eqs (\ref{30})
$F_{\mu\nu}F^{\mu\nu}=\frac{2Q_{1}^{2}}{\xi^{4}}$ and
$G_{\mu\nu}G^{\mu\nu}=\frac{2Q_{2}^{2}}{\xi^{4}}$, and we can
prove that Eqs (\ref{33}) are satisfied. The dilaton Eq.
(\ref{32}) takes the form
\begin{equation}
(\lambda^{2}\xi^{2}\psi')'=2(\frac{dh}{d\psi}Q_{1}^{2}+
\frac{dk}{d\psi}Q_{2}^{2})\xi^{-2}\label{35}
\end{equation}
where prime denotes differentiation with respect to $r$ as before.
The non-vanishing components of the Ricci tensor of the metric
(\ref{7}) are $R_{00}$, $R_{11}$, $R_{22}$ and
$R_{33}=sin^{2}\theta{R_{22}}$, and for the first three components
we get respectively from Eqs (\ref{34}) the relations
\begin{equation}
(\lambda^{2})'' +(\lambda^{2})'(\xi^{2})'\xi^{-2}=2(hQ_{1}^{2}+
kQ_{2}^{2})\xi^{-4} \label{36}
\end{equation}
\[-(\lambda^{2})''\lambda^{-2}-2(\xi^{2})''\xi^{-2}-
(\lambda^{2})'(\xi^{2})'\lambda^{-2}\xi^{-2}+
[(\xi^{2})']^{2}\xi^{-4}=(\psi')^{2}\]
\begin{equation}-2(hQ_{1}^{2}+
kQ_{2}^{2})\lambda^{-2}\xi^{-4} \label{37 }
\end{equation}
\begin{equation}
-[\lambda^{2}(\xi^{2})']'+2=2(hQ_{1}^{2}+ kQ_{2}^{2})\xi^{-2}
\label{38}
\end{equation}
Eqs (\ref{35})-(\ref{38}) form a system of four equations for the
three unknowns $\lambda^{2}$, $\xi^{2}$ and $\psi$. The
expressions (\ref{12}) and (\ref{13}) with $A$, $B$, $\alpha$ and
$\psi_{0}$ arbitrary constants, give again
a solution of this
system provided that the following relations are satisfied
\begin{equation}
g_{1}=\frac{AB(\sqrt{c^2+1}-c)}{2Q_{1}^{2}\sqrt{c^2+1}}e^{-(c+\sqrt{c^2+1})\psi_{0}}
\label{39}
\end{equation}
\begin{equation}
 g_{2}=\frac{(\alpha-A)(\alpha-B)(\sqrt{c^2+1}
+c)}{2Q_{2}^{2}\sqrt{c^2+1}}e^{-(c-\sqrt{c^2+1})\psi_{0}}
\label{40}
\end{equation}
Therefore in this model our solution is given by Eqs (\ref{13}),
(\ref{15}), (\ref{30}), (\ref{39}) and (\ref{40}). However in this
case $\psi_0$ is not really a parameter because of the symmetry
(\ref{31}). For example using this symmetry we can make
$\psi_0=0$. This means that, since the solution has the parameters
$A$, $B$, $\alpha$, $Q_1$ and $Q_2$, which must satisfy Eqs
(\ref{39}) and (\ref{40}), it has three free parameters. We can

take as such parameters the magnetic charges $Q_1$ and $Q_2$ and
the mass $M$, which is given by Eq.(\ref{17}). The conserved
dilaton charge $D$ of the solution is
\begin{equation}
D=\frac{1}{4\pi}\int_S J_\mu dS^\mu=-\frac{\alpha}{\sqrt{c^2+1}}
 \label{41}
\end{equation}
where the integral is over a two-sphere at spatial infinity. This
charge can be expressed in terms of the free parameters $Q_1$,
$Q_2$ and $M$. Thus according to Ref. \cite{Nu} the above solution
is not "hairy", and according to Ref. \cite{Kan} it has a hair of
secondary type.

I am very grateful to A. Kehagias for many illuminating
discussions and suggestions.

\end{document}